\documentclass[prb, twocolumn, showpacs]{revtex4}

\usepackage{graphicx}\usepackage{url}\usepackage{color}\usepackage{amssymb}
\begin{document}
\title{Near-isotropic performance of intrinsically anisotropic high-temperature superconducting tapes due to self-assembled nanostructures}

\author{Y. L. Zuev, D. K. Christen, S. H. Wee, A. Goyal, and S. W. Cook}

\affiliation{Oak Ridge National Laboratory, Oak Ridge, TN 37831, USA}


\date{\today}

\begin{abstract}
  High-temperature superconductors (HTS) are finding use as high - current wires for po\-ten\-tial ap\-pli\-ca\-tion in po\-wer de\-vi\-ces.  The dependence of their basal-plane critical current properties on orientation in a magnetic field can pose important design problems.  Here we report material and operating parameter conditions where prototype HTS tape conductors exhibit critical current characteristics that are essentially field-orientation independent.  The novel phenomenon is observed for specific magnetic field intensities that depend on the operating temperature, and in materials having strong flux pinning by extended nanoscale structures aligned roughly along the HTS crystalline $c$-axis.  The effect can be described by a simple model for the field dependence of critical current density, generalized for anisotropic electronic response.  This description may provide insight into means to fine-tune the material properties for nearly isotropic performance characteristics at a preferred field and temperature.

\end{abstract}
\maketitle 
\section*{Introduction}

The use of high-temperature superconductors (HTS) for a new generation of electrical conductors is nearing reality~\cite{Haught,Service}.  Ingenious research and development efforts have overcome many of the formidable obstacles to formation of the brittle, ceramic materials into flexible wires, and in upgrading their properties to levels that provide useful, nearly energy-loss free current conduction~\cite{Foltyn,Goyal1}.  The second-generation (2G) wires utilize the class of these layered, cuprate materials known by the chemical formula \textit{R}Ba$_{2}$Cu$_{3}$O$_{~7}$ (RBCO), where \textit{R} is a rare-earth, or mixture of rare earth ions, most typically Y, Nd, Sm, Gd, or Dy.  In their 2G usage, the HTS materials are deposited as an epitaxial coating on a near single-crystal template in the form of a long flat metallic-alloy tape.  The preference for the RBCO material class of HTS is due to its relatively smaller electronic anisotropy, which is intrinsic to the layered HTS crystal structure.  Aside from mechanical properties issues, many of the developmental problems have stemmed from this strongly anisotropic electronics and its ramifications for superconducting properties.  Indeed, the daunting task of achieving a near-single crystal coating over kilometer lengths is demanded by the fact that supercurrents can flow strongly only within the \textit{a-b }basal CuO$_{2}$-planes, compared to the relatively weaker conduction along the long, \textit{c-}axis of the crystalline orthorhombic structure.  In addition, it was necessary to achieve \textit{biaxial} alignment, since even small rotations of adjacent grains about the \textit{c }axis leads to diminished intergrain boundary currents (the weak-link problem), which likely arises from a combination of fundamental electronic and structural characteristics~\cite{Gurevich}.

At the basic level, the anisotropy has been described in terms of the effective supercurrent mass anisotropy ratio, characterized by the parameter $\gamma=\sqrt{m_c/m_{ab}}$.  For the RBCO materials, $\gamma=5\:-\:7$, compared to a value ~60 - 150 for the bismuth-based HTS (e.g., Bi$_{2}$Sr$_{2}$CaCu$_{2}$O$_{x}$ and Bi$_{2}$Sr$_{2}$Ca$_{2}$Cu$_{3}$O$_{x}$) used in the first-generation HTS wires.  The anisotropy is of practical importance because of its affect on the intragrain flux pinning - the immobilization of the quantized magnetic vortices that permeate superconductors in the presence of magnetic field, whose orientation with respect to the basal $ab$-plane can vary.  In this paper we are concerned exclusively with in-plane currents, and not at all with the currents, flowing along the $c$-axis. Electrical current density up to a critical value, $J_c$, is sustained only because the Lorentz-like driving force exerted by the current on the vortices is offset by the pinning force, thus preventing the energy dissipation that accompanies vortex motion.  For a given defect, the pinning efficacy is limited by current and thermally driven vortex distortions~\cite{Brandt}, which are determined by the vortex line tension for supercurrents flowing in basal planes, $\tilde{\epsilon}=\frac{\varphi_0H_{c1}}{4\pi\gamma^2}$.  Relative to that of an isotropic material, this value is smaller by the factor $1/\gamma^2$ for vortices along the \textit{c} axis, and leads to conditions of intrinsically depressed temperatures and fields where loss-free currents can be sustained, even for the case of optimal, columnar flux pinning defects with dimensions of a few nanometers that nearly match the vortex normal core~\cite{Figueras}.  Nevertheless, practical levels of performance can be achieved, as demonstrated by a large body of work on the modification, measurement, and analysis of materials with artificially introduced defect structures~\cite{Civale1}, and by recent advances in materials processing that invokes self-assembly of second-phase correlated nanostructures~\cite{Civale2,Goyal1,Kang,Yamada,Wee1,MacManus}.  

\begin{figure}[t]
\centering
\includegraphics[width=\columnwidth]{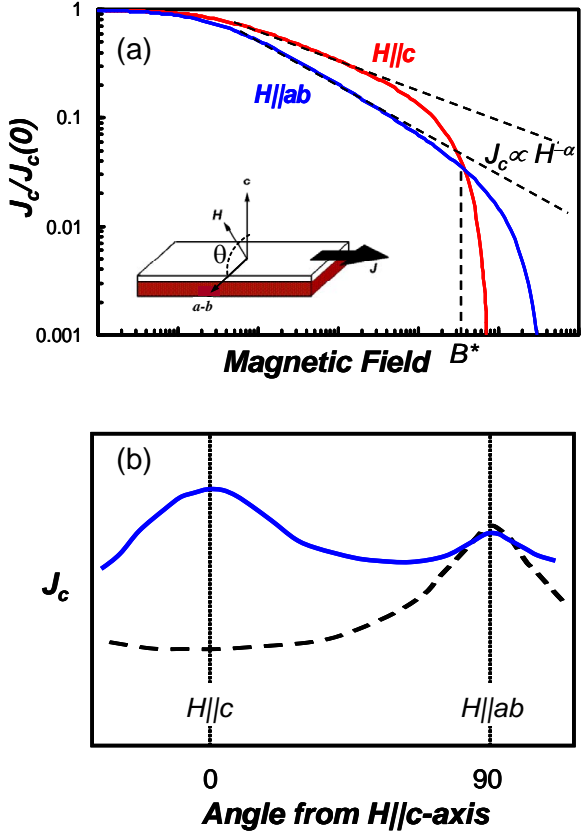}
\caption{(color online)(a) Example of the field dependent $J_c$ for the orientations $H\parallel c$ and $H\parallel ab$ as shown in the inset.  (b) Comparison of the orientation dependent $J_c$ at intermediate field.  Solid curve: $\alpha_c<\alpha_{ab}$ as in (a), the case for strong correlated pinning near the \textit{c}-axis.  Dashed curve: $\alpha_c\approx\alpha_{ab}$, the case for isotropic pins.}
\end{figure}

The net dependence of $J_c$ on the magnetic field magnitude and orientation with respect to the crystal structure depends on a complicated interplay between the pinning defect structure and the HTS electronic anisotropy.  Extended defects, such as lines or planes, lend toward enhancements in $J_c$ when the vortices are parallel, whereas isotropic, equiaxed pins should yield an angular dependence that depends on an effective magnetic field value $H_{\mathrm{eff}}=H/\gamma(\theta)$, where the external field $H$ is scaled by the angular-dependent mass anisotropy $\gamma(\theta)$~\cite{Blatter}
\begin{equation}
\gamma(\theta)=\left[\cos ^2(\theta)+\frac{m_{ab}}{m_c}\sin ^2(\theta)\right]^{-1/2}
\label{eq:gamma}
\end{equation}

Here, $\theta$ is the angle as the field is rotated away from the \textit{c }axis toward the \textit{a-b }planes, all for the case of maximum Lorentz force, with $\mathbf{J}$ in the \textit{a-b }planes, perpendicular to $\mathbf{H}$, as shown in Fig. 1(a). Typically, $J_{c}$ decreases with field magnitude as shown schematically in Fig. 1(a)~\cite{Foltyn,Aytug}.  At intermediate fields, isotropic pins might yield $J_c(\theta)$ as shown by the dashed line in Fig. 1(b).  Extended defects can produce a drastic enhancement in $J_c$, and strong deviations from this dependence.  The solid line in Fig. 1(b) illustrates effects that can be observed for columnar defects that are aligned near the \textit{c} axis~\cite{Goyal2}.  For these strong-pinning nanostructures, in some cases $J_{c,c}>J_{c,ab}$ over a range of intermediate fields~\cite{Wee2}, as also illustrated in Fig. 1(a).  In this field regime, it is often found that the field dependence is given by, $J_c\propto H^{-\alpha}$, where the power-law exponent $\alpha$ takes on values that are expected to be in the range 0.5 - 0.6, but can be smaller for very strong pinning, reflecting relatively better retention of $J_c$ with field.  Indeed, Fig. 1(a) illustrates a case where $\alpha_c < \alpha_{ab}$.  The rapid decay of $J_c$ toward zero at the irreversibility field $H_{\mathrm{irr}}$ is a consequence of depressed vortex line tension and strong thermal activation mentioned previously.  At finite temperature, $H_{\mathrm{irr}}$ is always less than the upper critical field $H_{c2}$ at a given orientation $\theta$, but in principle can be somewhat larger than the vortex lattice melting field $B_m$ (meaning that a vortex-matching array of optimal extended defects can provide some pinning of a vortex liquid)~\cite{Figueras,Blatter}. Since both $H_{c2}$and $B_m$ scale with $\gamma(\theta)$~\cite{Blatter}, we may expect $B_{\mathrm{irr},ab} > B_{\mathrm{irr},c}$, with a consequence that there will exist some crossover field $B^*$ at which $J_c$ is the same for both orientations, as illustrated in Fig. 1(a).  
\begin{figure}[t]
\centering
\includegraphics[width=\columnwidth]{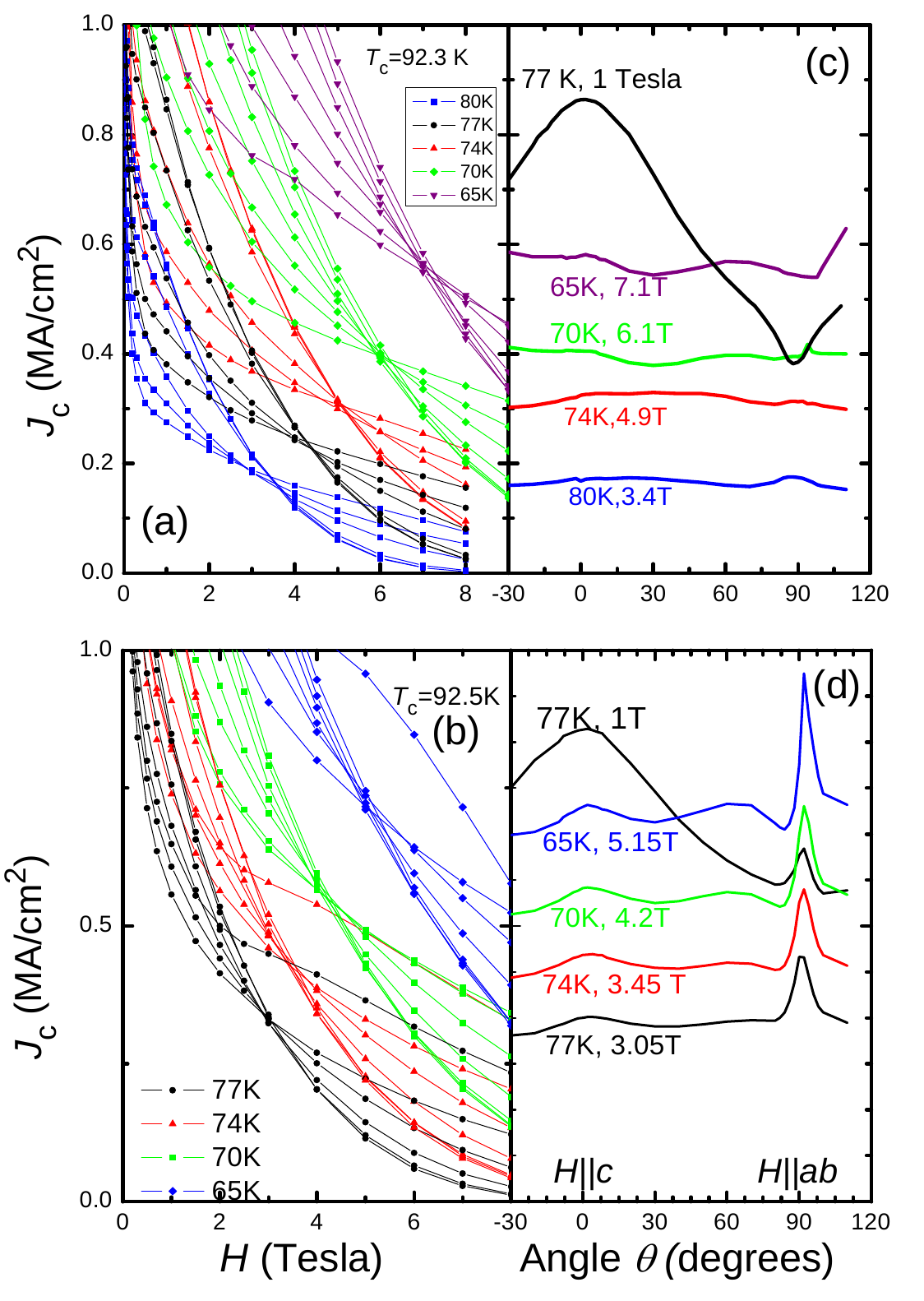}
\caption{(color online)(a,b) Critical current density as a function of applied magnetic field for samples "A" and "B" at different temperatures (shown by colors), and at sample orientations $H\parallel c$ to $H\parallel ab$.  At each temperature, the \textit{J}\textit{$_{c}$}(\textit{H}) isoclines for all field orientations cross at approximately the same field \textit{B*}(\textit{T}), with the exception of sample B for the special case $H\parallel ab$.  (c,d) The corresponding orientation dependence taken at the crossover fields \textit{B*}.  For each sample, the commonly used benchmark data of $J_c(\theta)$ at 77K, 1T are shown for comparison.}
\end{figure}

\section*{Experimental Results}

\begin{figure}[t]
\centering
\includegraphics[width=\columnwidth]{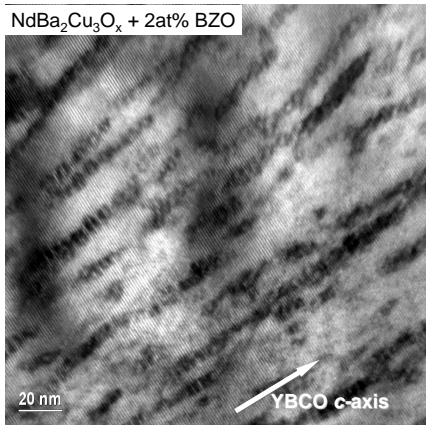}
\caption {Cross-section TEM image showing a columnar array of self-assembled stacked BaZrO3 precipitates.  The correlated defects are distributed nearly parallel to the NdBCO \textit{c} axis, and provide very strong vortex pinning.}
\end{figure}

We report here the surprising observation and phenomenological interpretation of a common field $B^*$ that exists for a wide range of orientations in the field, yielding an essentially constant, isotropic value of $J_c(B^*)$.  This phenomenon is illustrated in Fig 2, for two samples that have the characteristics of very strong pinning for fields aligned near the $c$-axis.  Here, $J_c$ is plotted as a function of magnetic field at several different temperatures and for orientations of the sample that span the range from $H\parallel c$ to $H\parallel ab$, as represented schematically by the inset of Fig. 1(a).  The transport critical current density $J_c$ is measured by the 4-terminal electrical technique in the full Lorentz-force configuration with the current also perpendicular to the applied field.  These particular samples are NdBa$_{2}$Cu$_{3}$O$_{x}$, ~0.8 $\mu$m thick, deposited by pulsed laser ablation on commercial tape templates from SuperPower, Inc.~\cite{Parans,SuperPower}. The films were deposited from a source target that contains ~2 vol\% BaZrO$_{3}$ (BZO) as a second phase.  Both samples have $T_C=92$ K. As reported previously~\cite{Civale2,Kang,Yamada,Wee1,MacManus}, appropriate growth conditions yield an array of BZO nanoparticles, stacked in columns aligned about the cuprate material \textit{c}-axis.  The resulting columnar structures, which may result from strain-driven self-assembly~\cite{Goyal1}, can have diameters in the range of 5 nm and inter-column spacings that correspond to matching vortex arrays of several Tesla.  Such nanostructures, as shown in the cross-section TEM image of Fig. 3, provide very strong flux pinning, characterized by particularly small values of $\alpha$ for aligned vortex orientations~\cite{Foltyn, Civale2}.

The sample "A" of Fig. 2(a) exhibits temperature dependent common crossover fields $B^*$ for all orientations, while the sample "B" of Fig. 2(b) shows convergence over a wide range of orientations, but enhanced $J_{c}$ for $H\parallel ab$.  Figures 2(c) and 2(d) show the angular dependence of $J_c$ taken near each $B^*$, and compared to the commonly used benchmark measurement at 77K and 1 Tesla.  For both samples "A" and "B" this latter standard case at 1 Tesla corresponds to a field that falls below $B^*(77K)$, and emphasizes the strong pinning for a broad angular range about $H\parallel c$.  Aside from the special situation $H\parallel ab$ of Fig. 2(d), the observed $J_c(B^*, \theta)$ values for both samples are constant to within a few percent.

\section*{Phenomenological Model Description}

We show now that this unique behavior can be described by generalization of a phenomenological model that has been used previously to parameterize the field dependence of $J_c$ for the case $H\parallel c$\cite{Aytug}.  The relationship is given by
\begin{equation}
\frac{J_c(B,\theta)}{J_c(0)}=\left[1+\frac{B}{B_0(\theta)}\right]^{-\alpha(\theta)}\left[1-\frac{B}{B_{\mathrm{irr}}(\theta)}\right]^2
\label{eq:model}
\end{equation}


In Eq. (2), the field $B_0$ sets the scale for transition from the low-field "$J_c$-plateau" of Fig. 1(a) to the power-law regime where $J_c\propto B^{-\alpha}$.  For a physical interpretation, $B_0$ might be regarded as the field above which inter-vortex interactions compete with vortex-pinning to suppress the critical current, while fields below $B_0$ characterize the individual vortex pinning regime ($J_c$ relatively field independent).  As seen below, for the present samples typical values of $B_0$ range from 5 to 10 mT, and cannot be determined with enough certainty to assign an unambiguous angular dependence.  The factor involving $B_{\mathrm{irr}}$ is largely phenomenological, with the exponent value of 2 determined to be empirically applicable~\cite{Aytug}.  However, by applying theoretical relationships for a special case of thermally-activated depinning from vortex-underfilled columnar defects~\cite{Brandt}, we have found that there is qualitative foundation for this form. 

\begin{figure*}[t]
\centering
\includegraphics[width=\textwidth]{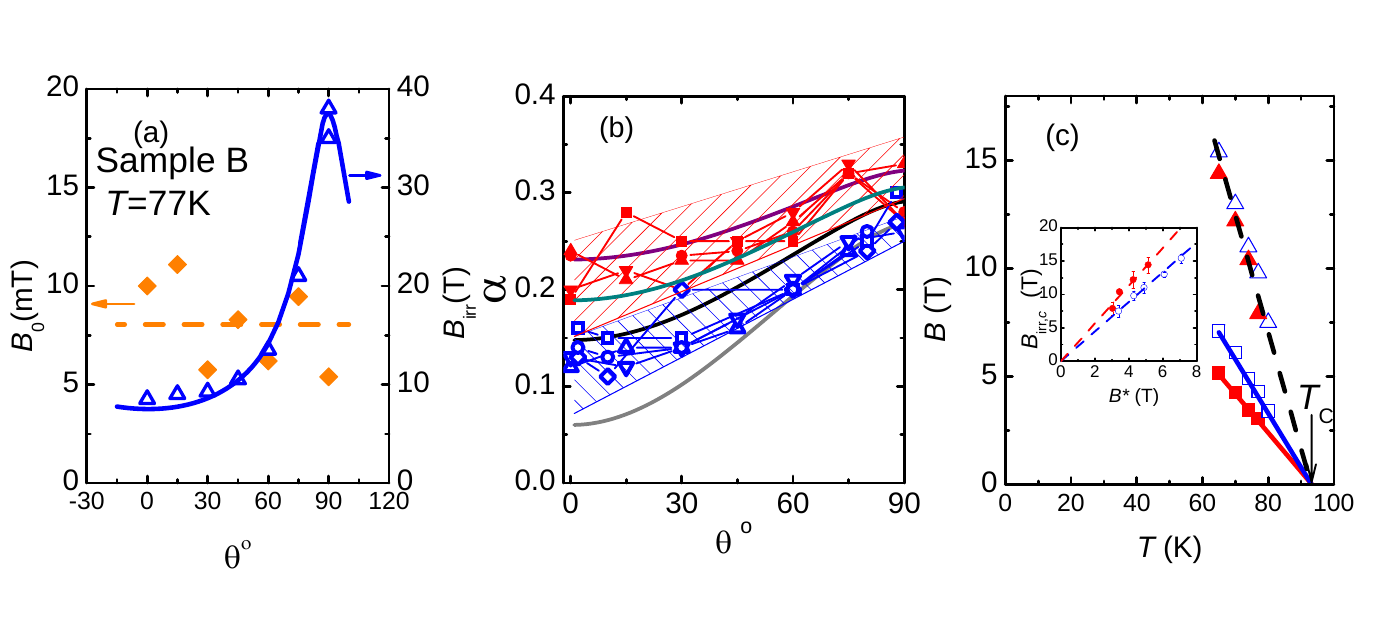}
\caption {(color online)(a) Values of the orientation dependent parameters $B_0$ and $B_{\mathrm{irr}}$ found from the modeling of Eq. (2) to the data of sample "B" at 77 K. The curve through the fitted $B_{\mathrm{irr}}(\theta)$ is the function $\gamma(\theta)$, Eq. (1), for an anisotropy ratio 5 and $B_{\mathrm{irr},c}=7.5$ T.  (b)Symbols: fitted values for the power-law exponent $\alpha$, for samples "A" (open symbols) and "B" (solid symbols) at T= 80K(\textcolor{blue}{$\lozenge$}), 77K(\textcolor{blue}{$\square$}\textcolor{red}{$\blacksquare$}), 74K(\textcolor{blue}{\Large $\circ$}\textcolor{red}{\Large $\bullet$}), 70K(\textcolor{blue}{$\triangle$}\textcolor{red}{$\blacktriangle$}) and 65K(\textcolor{blue}{$\triangledown$}\textcolor{red}{$\blacktriangledown$}). Values of $\alpha$ exhibit little systematic temperature dependence. Thick solid curves: $\alpha(\theta)$ that from the model would produce perfect convergence at $B^*=2$ T (purple), 2.5 T (green), 3 T (black) and 4 T (grey). The downturn of $\alpha$ at $H\parallel ab$ reflects the related upturn in $J_{c,ab}$.  As a good approximation for $\alpha(\theta)$ the linear bands may be used to gain insight. Hatched areas represent a spread in deduced values of $\alpha$ (c) The irreversibility field $B_{\mathrm{irr},c}$ (triangles) and crossover field $B^*$ (squares) for samples "A" (blue) and "B" (red). The inset shows that $B^*$ varies linearly with $B_{\mathrm{irr},c}$} 
\end{figure*}

As shown in Eq. (2), the three parameters $B_0$, $\alpha$, and $B_{\mathrm{irr}}$ can be orientation dependent.  As explained above, anisotropy of $B_{\mathrm{irr}}$ is anticipated on fundamental grounds.  However, a clear understanding of both the magnitude and angular dependence of $B_0$ and $\alpha$ are still lacking.  In any case, Eq. (2) provides a good description of the data at various orientations. Since $J_c(0)$ is directly measureable, the fits to Eq. (2) yield the three parameters $B_0$, $\alpha$, and $B_{\mathrm{irr}}$ at each angle $\theta$ and temperature $T$.  

Results for some deduced parameters are shown in Fig. 4.  Figure 4(a) is an example of the output for the characteristic fields $B_0$ and $B_{\mathrm{irr}}$ at 77 K.  While we cannot ascribe an angle dependence to $B_0$ due to uncertainties in, and insensitivity to, its values, the $B_{\mathrm{irr}}$ are well described by the solid curve, given by Eq. (1), for an intrinsic supercarrier mass anisotropy $\sqrt{m_c/m_{ab}}=5$
, a reasonable value for NdBCO.  Blue and red symbols in the figure 4(b) are the fitted power-law exponent $\alpha$ found for samples "A" and "B" of Fig. 2, respectively. Here different symbol shapes corespond to different temperatures, as explained in the figure caption. Aside from the special case $H\parallel ab$ in sample "B", it can be seen that $\alpha$ varies monotonically with angle.  Within the applicability of Eq.(2), we may solve for the "ideal" $\alpha(\theta)$ that would generate \textit{perfect} convergence at $B^*$ (i.e., $J_c(B^*,\theta)$ constant).  This yields the relation, 
\begin{equation}
\alpha(\theta)=\frac{\ln\left[\left(1-B^*/B_{\mathrm{irr}}(\theta)\right)^2/J^*\right]}{\ln\left(1+B^*/B_0\right)}
\label{eq:alpha}
\end{equation}

Equation (3) is plotted as the family of thick solid curves for different selected $B^* =$ 2, 2.5, 3 and 4 Tesla.  Here, $J^*=J_c(B^*)/J_c(0)$
 is fixed at the value 0.15, $B_0=8$ mT, and $B_{\mathrm{irr},c}=7.5$ T.  Due to the weak logarithmic sensitivity to angle through $B_{\mathrm{irr}}(\theta)$, we find that the relatively smooth, monotonic variation of this ideal $\alpha(\theta)$ can be approximated as linear, resulting in only minor smearing of a simulated model-calculated crossover point at $B^*$, basically very similar to that seen in the experimental data.

\section*{Temperature Dependence}


\begin{figure}[t]
\centering
\includegraphics[width=\columnwidth]{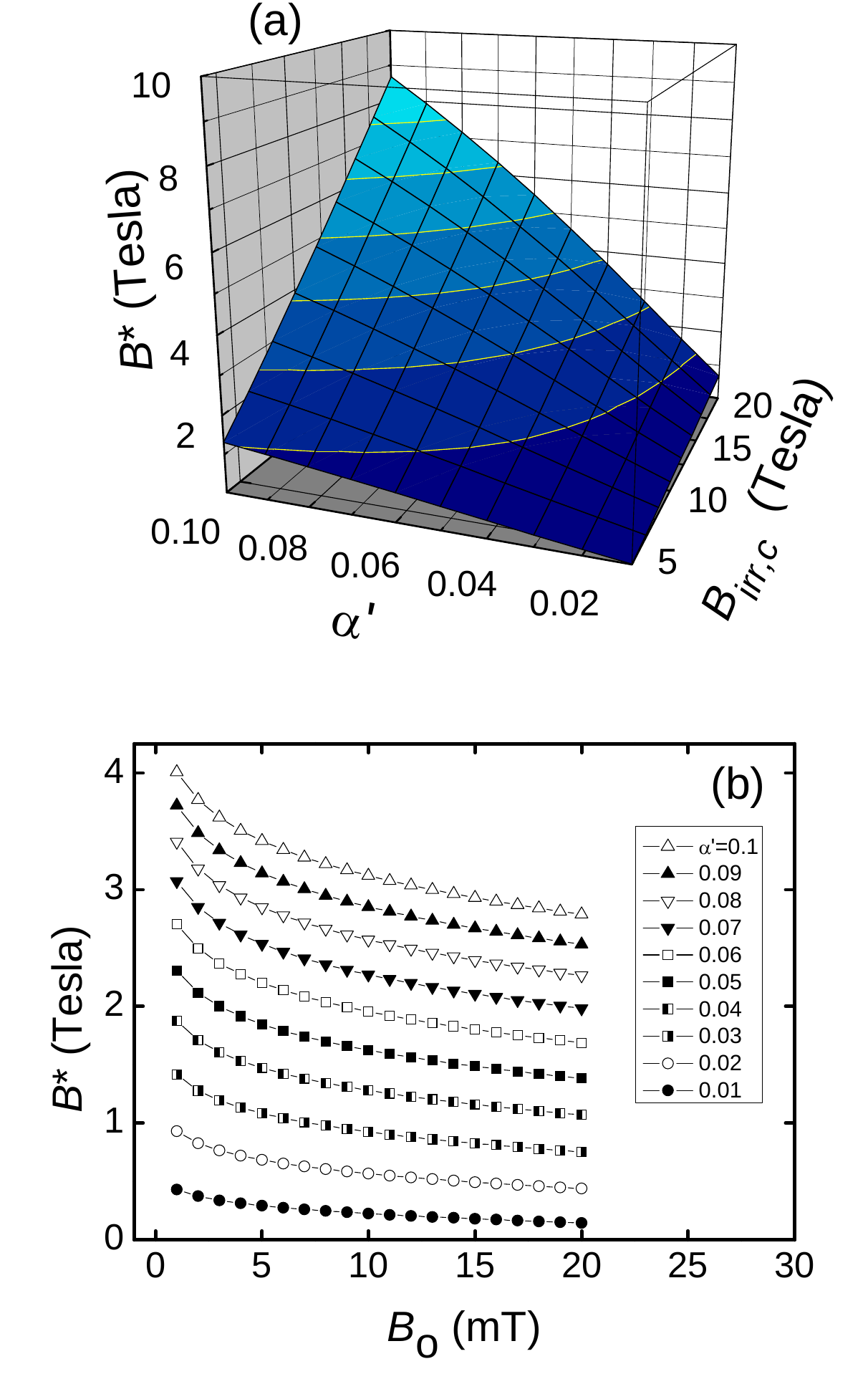}
\caption {(color online)(a) Dependence of $B^*$ on $B_{\mathrm{irr},c}$ and $\alpha^{\prime}$. Cuts along the $B_{\mathrm{irr},c}$ direction produce nearly straight lines, in accordance with fig. 4(c) inset. Cuts along the $\alpha^{\prime}$ direction produce a slight downward curvature (b) Dependence of $B^*$ on $B_0$ is relatively weak.}
\end{figure}

Figures 4(b) and (c) also show the temperature dependence of the relevant parameters in the investigated liquid nitrogen range (65 K - 77 K).  For the two samples, the extracted irreversibility fields at $H\parallel c$ are found to scale with their respective directly observable crossover fields $B^*$, while each vary nearly linearly with temperature, as shown in Fig. 4(c).  In contrast, within the scatter there is little systematic temperature dependence in $\alpha(\theta)$, and the hatched bands of Fig. 4(b) indicate that  linear approximations of the fitted values are reasonable.  However, it should be noted that the fitted values of $\alpha$ for sample "B" are less reliable due to systematic high-field deviations from the model.  As such, the magnitudes of $\alpha$ depend on the exact choice of the fitting range.  This problem is most likely due to "field-matching vortex pins" but does not affect the overall orientation dependence of $\alpha$.  Details will be reported elsewhere.  In working with only the first derivative of Eq. (3), one can gain insight into the predicted inter-relationships among the three parameters, since then the actual value of $J_c$ is not needed.  For example, in this case one might deduce an average slope by measurement at only two relevant orientations, e.g.,
\begin{eqnarray}
\alpha^{\prime}\equiv\left<\frac{d\alpha(\theta)}{d\theta}\right>=\frac{2}{\pi}(\alpha_{ab}-\alpha_c) =\\ \nonumber
\frac{4}{\pi}\frac{\ln\left[1-\sqrt{\frac{m_{ab}}{m_c}}\frac{B^*}{B_{\mathrm{irr},c}}\right]-\ln\left[1-\frac{B^*}{B_{\mathrm{irr},c}}\right]}{\ln(1+B^*/B_0)}
\label{eq:alphaprime}
\end{eqnarray}

From Eq. (4) we can deduce the model dependence of $B^*$ on $B_{\mathrm{irr}}$ and $\alpha^{\prime}$ and the surface of Fig. 5(a) demonstrates the predicted behavior.  Despite the apparently non-transparent form of Eq. (4), it generates a nearly linear dependence of $B^*$ on $B_{\mathrm{irr},c}$, consistent with the experimental findings of Fig. 4(c).  Figure 5(b) confirms that the predicted crossover field is not strongly dependent on $B_0$ (which we have taken as constant in the above analysis).

\section*{Discussion and Conclusions}

At this point, some general statements regarding the overall experimental behavior, taken in the context of the model description, Eq. (2), must be made:

\begin{itemize}

\item

The crossover phenomenon occurs at the field $B^*$ yielding $J_c(B*,\theta)$ constant, only for a situation where the flux pinning at intermediate fields is increasing as the field orientation approaches the \textit{c} axis (i.e. $d\alpha(\theta)/d\theta>0$), and the irreversibility field increases as orientation approaches \textit{ab} (consistent with the HTS electronic mass anisotropy)
\item

The crossover effect is robust, provided $\alpha(\theta)$ is monotonic and smoothly varying with orientation.
\item

For an approximate linear variation of $\alpha(\theta)$, the model yields an only slightly smeared crossover point, and scaling of the crossover field $B^*$ with the irreversibility field $B_{\mathrm{irr}}$.
\item

The crossover field $B^*$ does not depend on the value of $\alpha$, but only on its (average) rate of variation with orientation, $\alpha\prime$ (Eq. (4)), whereas the magnitude of the isotropic $J_c(B^*)$ is enhanced for smaller values of $\alpha$ (and, of course, large $B_{\mathrm{irr}}$) (Fig 1(a)).

\end{itemize}

\begin{figure}[b]
\centering
\includegraphics[width=\columnwidth]{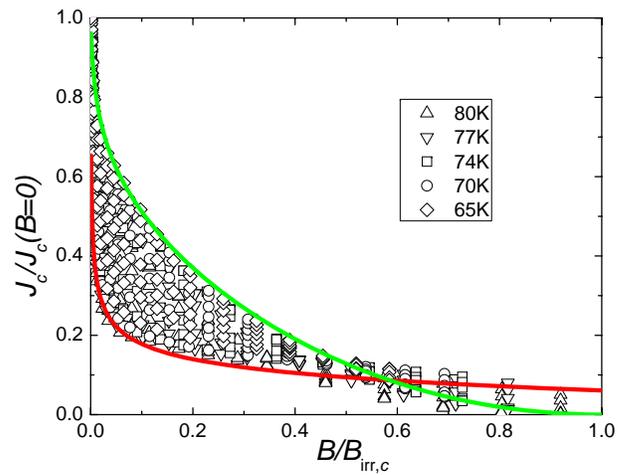}
\caption {(color online) $J_c(\theta)$ at different temperatures, scaled by the irreversibility field $B_{\mathrm{irr},c}$ and by each self-field value of the critical current density $J_c(0,T)$.}
\end{figure}

The present study has been conducted over a limited temperature range for practical reasons (limitation of magnetic field and electric current levels).  However, provided Eq. (2) is descriptive to lower temperatures, and if $\alpha$ remains nearly temperature independent, then the phenomenon can be described by a set of universal curves given in terms of scaled variables, such as illustrated for the present data in Fig. 6.  Similar scaling with angle has been reported, but in that case for a different type of pinning system for which $\alpha$ is nearly angle-independent (i.e., the limit $\alpha^{\prime}\rightarrow 0$).  In that case, the scaling maps the data onto a single curve~\cite{Civale3}, analogous to the large body of work that has been done for pinning models of isotropic low-temperature superconductors, where scaling with respect to temperature has been obtained for the bulk pinning force density $F_p$, with forms such as $F_p=J_cB\approx b^p(1-b)^q$
, but where $b=B/B_{c2}$\cite{DewHughes}.  As mentioned above, recent advances in deposition of HTS films under conditions that purposely produce self-assembled extended or strong-pinning defect structures have yielded the very strong \textit{c-}axis performance needed for the present phenomenon~\cite{Civale2,Kang,Yamada,Wee1,MacManus}.  Although nearly isotropic critical currents have been reported or predicted under certain or special conditions~\cite{Solovyov,Rodriguez, Long}, there has been no previous recognition of the nearly universal crossover behavior, which should occur and yield isotropic $J_c$ only in this class of especially strong-pinning materials.  

From a practical viewpoint, such a scaling relationship could help guide design of HTS wires for targeted applications parameters, since there is a need for isotropic performance characteristics in devices operating in substantial magnetic fields, such as motors, generators, levitation magnets, etc.  Simulations from the model can offer predictive value in materials tailoring, provided processing conditions can be controlled to tune the physical parameters, most importantly $\alpha$ and $J_c(0)$.  In approaching selectivity of properties, it is unfortunate that there is little theoretical guidance in understanding the beneficial small $\alpha$ values (typically $< 1/3$) that can be realized for these particular self-assembled pinning nanostructures.  Previous theory has explored mechanisms that predict $J_c \propto 1/B^{\alpha}$, but which for particular defect types have yielded calculated power-law exponents of 1/2 (correlated extended)~\cite{Nelson} and 5/8 (large random)~\cite{Ovchinnikov,vdBeek}. Indeed, $\alpha$ values in this range have been reported for a variety of more weakly pinned HTS systems~\cite{Aytug,Civale3,Ijaduola,Dam,Klaassen}

In conclusion, we have reported the systematic observation and analysis of isotropic critical currents in HTS materials.  Interestingly, the phenomenon arises due to the combination of intrinsic electronic anisotropy coupled with the very strong vortex confinement due to arrays of self-assembled stacks of nanophase BaZrO$_{3}$, which are distributed with alignment near the \textit{c} axis in NdBaCuO films.  Nearly complete isotropy of $J_c$ is realized at the field $B^*(T)$, which is found to have the same temperature dependence as the orientation-dependent irreversibility field $B_{\mathrm{irr}}(T,\theta)$.  By generalizing a simple flux pinning model for angle-dependent parameters, the features of the observations can be described analytically, with potential predictive utility in the design of conductors for particular applications.  As mentioned above, the model Eq. (2) has limitations; for example, in the case of samples that show evidence for matching of the vortex density with that of the pinning array.  In that case, additional $J_c$ above that predicted by Eq. (2) is found at the field where every vortex could be accommodated by an extended defect.  Moreover, it is also possible that vortex alignment with correlated pins causes deviations of $H_{\mathrm{irr}}(\theta)$ from the descriptive effective mass anisotropy of Eq. (1)~\cite{Baily}.  Other subtle effects also have been observed and will be reported elsewhere.  This work places further perspective on and context to the challenges presently pursued by researchers and developers worldwide: to control, modify, and understand the effects of such special defect structures, which may enable optimization of $B^*$ and $J_c(B^*)$.

The authors are grateful to J. R. Thompson and D. F. Lee for useful discussions.  Y. L. Zuev and S. H. Wee would like to thank Oak Ridge Associated Universities for a postdoctoral fellowship.  This work was sponsored by the Department of Energy Office of Electricity Delivery and Energy Reliability (OE) -- Superconductivity for Electric Systems, and the Office of Basic Energy Sciences -- Division of Materials Sciences and Engineering. This research was performed at the Oak Ridge National Laboratory, managed by UT-Battelle, LLC for the USDOE.

\end{document}